\documentclass{JAC2003}

\usepackage{graphicx}

\begin{document}

\def\ep {$e^{+}e^{-}$ } 
\def\ee {$e^{-}e^{-}$ } 
\def\twentymrad {$20\; {\rm mrad}$ }
\def\twomrad {$2\; {\rm mrad}$ }

\title{\vspace*{-1.8cm}
\begin{flushright}
{\bf\large LAL/RT 06-08}\\
\vspace*{-0.1cm}
{\normalsize\bf EUROTeV-Report-2006-067}\\
\vspace*{-0.1cm}
{\normalsize\bf CARE/ELAN 2006-007}\\
\vspace*{-0.1cm}
{\large June 2006}
\end{flushright}
\vspace*{1,5cm}
OPTIMIZATION OF THE \ee OPTION FOR THE ILC\thanks{Work supported in part 
by FP5 (PROBE FOR NEW PHYSICS) contract RTN2-2001-00450, FP6 (CARE) contract 
RII3-CT-2003-506395 and (EUROTEV) contract RIDS-011899}}

\author{{\bf\large M.~Alabau Pons (IFIC-Valencia), P. Bambade, O. Dadoun}\\ 
{\it Laboratoire de l'Acc\'el\'erateur Lin\'eaire,}\\ 
{\it IN2P3-CNRS et Universit\'e de Paris-Sud 11, BP 34, 91898 Orsay cedex, France\vspace*{0,3cm}}\\
{\bf\large R. Appleby }\\ 
{\it The Cockcroft Institute and the University of Manchester,}\\ 
{\it Oxford Road, Manchester, M13 9PL, England
\vspace*{0,3cm}}\\
{\bf\large  A.~Faus-Golfe}\\
{\it IFIC,}\\
{\it Ed. Institutos de Investigaci\'on, 4607 Paterna (Valencia), Spain}
\vspace*{5cm}
}

\maketitle

\begin{abstract}
The \ee running mode is one of the interesting physics options at the 
International Linear Collider (ILC). The luminosity for \ee collisions is 
reduced by the beam-beam effects. The resulting beamstrahlung energy loss and 
beam-beam deflection angles as function of the vertical transverse offset are 
different compared to the \ep collisions. In this paper, the dependence of these 
observables with the offset for different beam sizes has been analyzed to 
optimize performances for the \ee mode, taking into account the requirements of 
the beam-beam deflection based intra-train feedback system. A first study of the 
implications for the final focus and extraction line optics is also presented 
for the cases of the 20 mrad and 2 mrad ILC base line crossing angle geometries.
\end{abstract}

\section{BEAM-BEAM EFFECTS}

\begin{figure}[htb]
\centering
\includegraphics*[width=85mm]{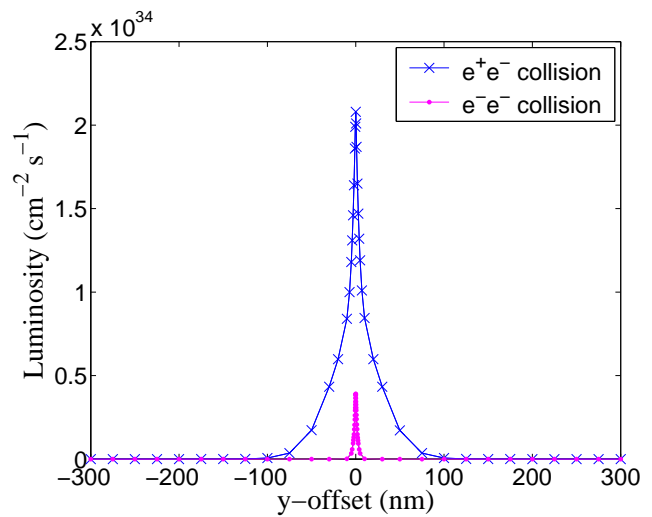}
\caption{Luminosity versus vertical half beam-beam offset, for \ep and \ee 
collisions simulated with GUINEA-PIG \cite{guinea}, using idealised Gaussian 
beam distributions with ILC nominal parameters at 500 GeV in the center-of-
mass.}
\label{lumi-GP}
\end{figure}

\begin{figure}[htb]
\centering
\includegraphics*[width=85mm]{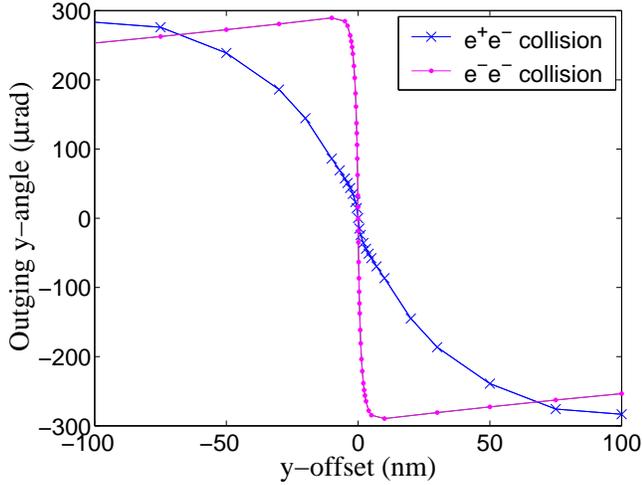}
\caption{Vertical deflection angle versus vertical half beam-beam offset, for 
\ep and \ee collisions at the ILC with nominal parameters at 500 GeV in the 
center-of-mass.}
\label{deflection-GP}
\end{figure}

At the Interaction Point (IP) of the ILC, beam-beam effects due to the strong 
electromagnetic fields that the bunches experience during collisions cause a 
mutual focusing, called pinch effect, which enhances the luminosity in the case 
of \ep collisions. The opposite is true for \ee collisions. In this case the 
luminosity is reduced by mutual defocusing, or anti-pinching and is only about 
20$\%$ of the \ep one (see Fig.~\ref{lumi-GP}). Moreover this repulsion between 
the bunches causes the luminosity to drop with a vertical offset at the IP much 
more rapidly for the \ee case than for \ep. Another effect of this strong 
repulsive electromagnetic field is the much steeper beam-beam deflection curve 
(see Fig.~\ref{deflection-GP}). Since the fast intra-train feedback system used 
to maintain the beams in collision at the IP \cite{feedback-GWhite} exploits 
these deflections as its main signal and because of the higher sensitivity to 
the vertical offsets, it is important to compare average performances for \ee 
and \ep for a set of representative values of initial beam offsets and bunch-to-
bunch jitter.

%%%%%%%%%%%%%%%%%%%%% Fig {lumi-GP}

%%%%%%%%%%%%%%%%%%%%% Fig {deflection-GP}

\section{FEEDBACK SIMULATION}
A simplified simulation of the feedback has been carried out using parametrized 
information from the last bunch crossing with a single proportional factor to 
relate the measured deflection angle of the outgoing beam to the correction of 
the offset at the IP \cite{feedback-Schreiber}. At frequencies of a few Hz 
corresponding to the ILC train repetition rate, offsets of order of hundreds of 
nm are predicted (see e.g. \cite{ground-motion}). In addition bunch-to-bunch 
jitter of a fraction of the beam size can be expected. The simulation has been 
done for different assumptions on the initial train offset and bunch-to-bunch 
jitter, and including a 10$\%$ error on the correction to represent the measured 
uncertainties. The factor relating the correction to the measured deflection 
angle was optimized, independently for \ep and \ee beam parameters, to maximize 
the speed of the correction without amplifying the bunch-to-bunch jitter by 
over-correcting. Nominal beam parameters \cite{parameters} were used at 
$\sqrt{s}=$ 500~GeV. The average luminosity loss over a train was found to be 
almost independent of the offset at the beginning of the pulse for the range 
considered (up to 500~nm). The \ee luminosity loss was however found to be a 
factor 2 greater compared to \ep for the same assumption on the jitter.This is 
due to the greater sensitivity to the vertical offset. The ability to decrease 
this sensitivity with alternative beam parameters could be important if jitter 
conditions are worse than expected, e.g. during early ILC operation.

\begin{figure}[htb]
\centering
\includegraphics*[width=85mm]{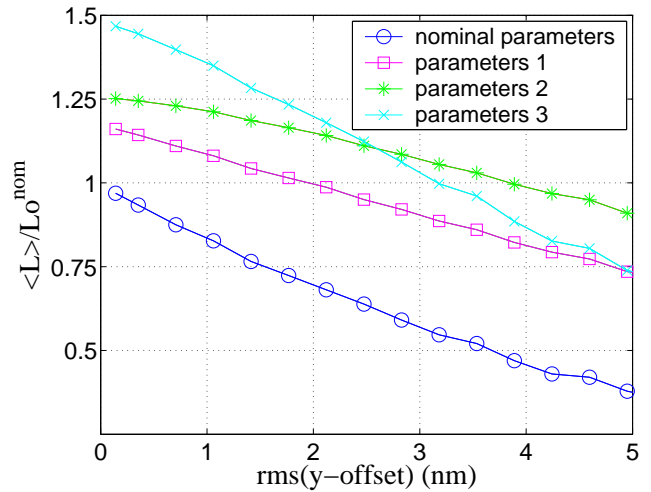}
\caption{Average train luminosity normalized to the peak luminosity with nominal 
parameters for \ee versus r.m.s. vertical offset difference between the beams. 
The results shown include a 100~nm initial offset.}
\label{feedback_eminus}
\end{figure}

With this purpose, sets of alternative beam parameters with smaller disruption 
have been derived by decreasing the bunch length and varying the transverse beam 
sizes, in order to maximize the luminosity while limiting the beamstrahlung 
energy loss to 5$\%$ (see Table~\ref{parameters_table}). These alternative 
parameters have increased luminosity, and some of them smaller sensitivity to 
the IP offset, compared to those obtained for the nominal case for \ee. The 
average train luminosity for different amplitudes of the jitter applied to each 
beam, is also improved compared to the nominal parameters (see 
Fig.\ref{feedback_eminus}).

%%%%%%%%%%%%%%%%%%% Fig {feedback_eminus}

\begin{table}[hbt]
\begin{center}
\caption{Luminosity and beamstrahlung energy loss for \ee collision for 
different parameter sets with a beam energy of 250~GeV. The nominal values for 
the beam sizes at the IP are $\sigma_{z0}^{*}=$300~$\mu$m and 
$\sigma_{x0/y0}^{*}=$655.2/5.7~nm and the nominal intensity is 
$N_{0}=2\times10^{10}$ particles.}
\vspace*{0,6cm}
\begin{tabular}{|l||c|c|c|c|c|}
\hline
\textbf{} &  \textbf{nom.} & \textbf{set 1} & \textbf{set 2} & \textbf{set 3} & 
\textbf{low P} \\ \hline \hline
$N/N_{0}$                       &    1    &     1        &     1        &      1       
&   0.5    \\ \hline
$\sigma_{z}^{*}$/$\sigma_{z0}^{*}$      &    1    &     0.7      &     0.5      
&      0.5     &   0.5    \\
$\sigma_{x}^{*}$/$\sigma_{x0}^{*}$      &    1    &     0.7      &     0.8      
&      0.9     &   0.7    \\
$\sigma_{y}^{*}$/$\sigma_{y0}^{*}$      &    1    &     1.5      &     1.5      
&      1       &   0.6    \\ \hline \hline
$\epsilon_{x}^{*}$($\mu$m)              &   10    &     10       &     10       
&     10       &   9.6    \\ 
$\epsilon_{y}^{*}$($\mu$m)              &    0.04 &      0.04    &      0.04    
&      0.04    &   0.03   \\ \hline 
$\beta_{x}^{*}$(mm)                     &   21.0  &     10.3     &     13.4     
&     17.0     &  10.0    \\ 
$\beta_{y}^{*}$(mm)                     &    0.4  &      0.9     &      0.9     
&      0.4     &   0.2    \\ \hline 
$L~(\times10^{33})$                     &    3.9  &     4.6      &     4.9      
&      5.8     &   3.0    \\
($cm^{-2}s^{-1}$)                       &         &              &              
&              &          \\ \hline
$\delta_{B}$ ($\%$)                     &    2.24 &     4.9      &     5.0      
&      4.3     &   2.2    \\ \hline
\end{tabular}
\label{parameters_table}
\end{center}
\end{table}

An additional set of parameters is also being investigated with only half of the 
bunch charge, while keeping the same number of bunches per train. It has a 
smaller peak luminosity (see Table~\ref{parameters_table}) and similar 
sensitivity to IP offsets as the \ee nominal case. Such a parameter set could be 
important for early ILC operation and flexibility.

\section{OPTICS STUDIES FOR THE \twentymrad CROSSING-ANGLE GEOMETRY}
\subsection{Final focus}
\begin{figure}[h]
%\centering
\hspace*{-1cm}
\rotatebox{-90}{\includegraphics*[width=65mm]{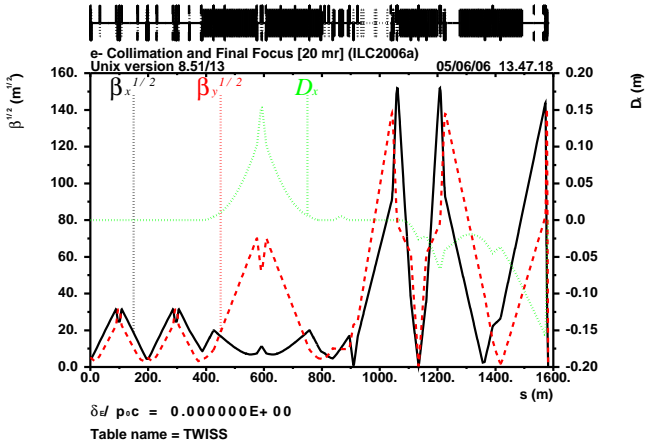}}
\caption{Optics solution for the parameter set 2, for the 20 mrad crossing-angle 
geometry.}
\label{20mrad-par2}
\end{figure}

The optics of the Final Focus (FF) (corresponding to the 20~mrad crossing-angle 
geometry) has been refitted to obtain the new $\beta$-functions at the IP for 
the alternative beam parameters in Table~\ref{parameters_table}. Only the 
quadrupoles upstream of the chromatic correction section and the sextupoles were 
readjusted. This allows to maintain the geometry and overall optimization of 
high order effects. The $\beta$-functions and the dispersion for the parameter 
set 2 in Table~\ref{parameters_table} are shown in Fig.~\ref{20mrad-par2}.

%%%%%%%%%%%%%%%%%%%% Fig {20mrad-par2}

\begin{figure}[htb]
\centering
\includegraphics*[width=79mm]{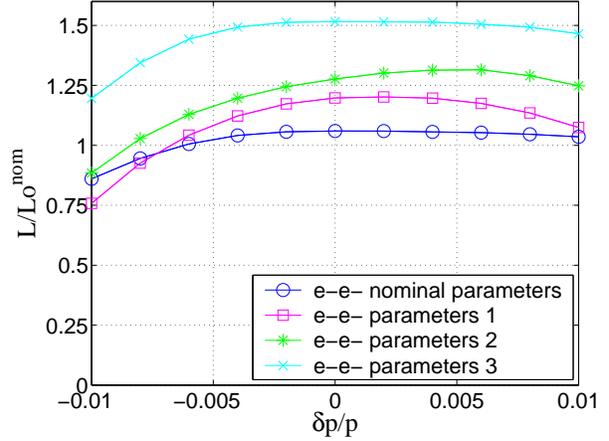}
\caption{Optical bandwidth for the different \ee set of parameters. All the 
luminosities are normalized with respect to that obtained with GUINEA-PIG at 
nominal energy for nominal parameters and with ideal beam parameters at the IP 
(i.e. without higher order optical effects).}
\label{bandwidth-20mrad}
\end{figure}

The optical bandwidth for the different sets of parameters has been studied, 
considering beams with a uniform flat momentum distribution with an energy 
spread of 0.1$\%$. The distribution of particles at the entrance of the FF, was 
created with PLACET \cite{placet} for different average momentum offsets. The 
beam was then tracked through the FF with MAD8 \cite{mad} and used as input for 
GUINEA-PIG to compute the luminosity.  The results (see Fig.~\ref{bandwidth-20mrad}) show similar off-momentum behavior for all parameter sets, with the 
alternative sets having better peak performance.

%%%%%%%%%%%%%%%%%%%% Fig {bandwidth_20mrad}

\subsection{Extraction line}
\begin{figure}[htb]
%\centering
\hspace{-1cm}
\rotatebox{-90}{\includegraphics*[width=65mm]{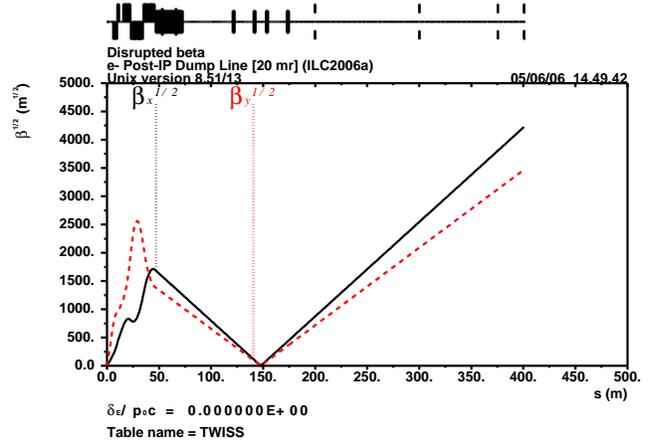}}
\caption{$\beta$-functions for the disrupted outgoing beam for the parameters 
set 2.}
\label{extraction-optics-20mrad}
\end{figure}

\begin{figure}[htb]
\centering
\includegraphics*[width=84mm]{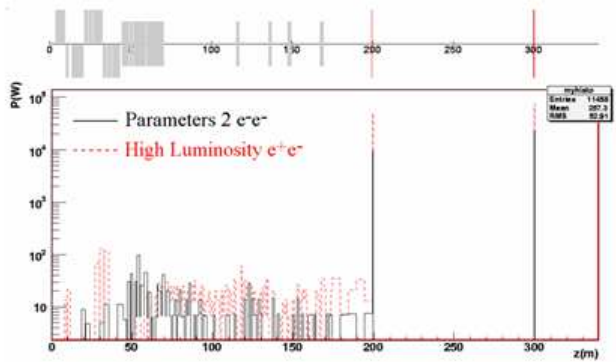}
\caption{Power losses along the extraction line for the parameter set 2 for \ee 
and the high luminosity parameters, for \ep at 500 GeV in the center-of-mass.}
\label{extraction-tracking-20mrad}
\end{figure}

The effective parameters corresponding to the disrupted beam have been computed 
along the extraction line for the different beam parameter sets in Table 
\ref{parameters_table} (see Fig.~\ref{extraction-optics-20mrad}), and for the 
different parameters sets suggested for \ep in \cite{parameters}. The largest 
values found for the $\beta$-functions in the \ee and \ep cases were comparable. 
The tracking of the disrupted beams has been simulated with BDSIM \cite{bdsim} 
and the power losses along the line have been computed. For the parameter set 2 
the losses are smaller than for the high luminosity parameters for \ep (see 
Fig.~\ref{extraction-tracking-20mrad}).

%%%%%%%%%%%%%%%%%% Fig {extraction-optics-20mrad}

%%%%%%%%%%%%%%%%%% Fig {extraction-tracking-20mrad}

\section{OPTICS STUDIES FOR THE \twomrad CROSSING-ANGLE GEOMETRY}
In the 2~mrad crossing-angle geometry the spent beam is transported off-axis 
through the last defocussing quadrupole of the final focus. The kick produced by 
this quadrupole is used to extract the spent beam. This scheme doesn't work for 
the \ee option unless one can reverse the signs of the focusing and defocusing 
final doublet quadrupoles (and sextupoles), while keeping at least the strength 
of the last quadrupole to maintain the kick needed for extraction. A first 
attempt in this direction \cite{snowmass-Seryi} indicated that this was 
feasible, but large $\beta_{y}$-value had to be used at the IP to limit the 
vertical beam size in the final doublet, which is important to keep reasonable 
collimation depth. This resulted however in about a factor 2 lower peak 
luminosity. Improvements with half the bunch length are also being investigated, 
with for example $\beta^{*}_{x/y}=10/3$ mm. In this case, more acceptable 
overall performance is expected.

 \vspace*{0,5cm}
\section{CONCLUSIONS AND PROSPECTS}
For the 20 mrad crossing-angle geometry, beam parameters can be obtained for the 
\ee option by decreasing the bunch length, with improved peak luminosity, 
smaller sensitivity to IP offsets, and similar beam losses in the extraction 
line as those found for \ep. For the 2 mrad crossing-angle, it is necessary to 
go to larger vertical beam sizes at the IP, which decreases the luminosity. With 
half of the bunch length and optimizing the transverse beam sizes taking into 
account collimation requirements, a first study indicates that some of this 
reduction can be recovered. In the near future, these problems will be studied 
to further characterize the \ee option at the ILC.

\vspace*{0,5cm}
%%%%%%%%%%%%%%%%%%%% BIBLIOGRAPHY %%%%%%%%%%%%%%%%%%%%%%%%%

\end{document}